# Dynamical phase diagram of parity-time symmetry with competing saturable channels

Sunkyu Yu, Xianji Piao, and Namkyoo Park*

*Photonic Systems Laboratory, Department of Electrical and Computer Engineering, Seoul National University, Seoul 08826, Korea*

*Correspondence to*: nkpark@snu.ac.kr

**Nonlinear channels play a critical role in realizing dynamical functions. Neural ionic channels[1] and non-volatile memristors[2] each derive representative biological and electrical functionalities, such as repetitive firing[3] or pinched hysteresis[2]. In electromagnetics, saturable channels of amplification[4] or absorption[5] provide a large nonlinearity for nonequilibrium wave dynamics, from conventional lasing[6] to mode locking[7,8] to recent achievements of the non-reciprocity[9-13] in complex potentials. Here, we investigate the dynamical phase diagram of parity-time symmetric systems, governed by competing nonlinear channels of saturable amplification and absorption. Determined by the relative strength and saturation level of the channels, three distinctive phases of fast- and slow-response equilibriums, and an oscillating nonequilibrium are demonstrated. On phase boundaries, we also reveal the chaotic existence of the strong oscillation state, which allows the non-reciprocal realization of repetitive resonator firing with fully tunable time delays. This work will promote the wave-based realization of nonlinear and chaotic temporal functions, toward light-based neural systems.**

Interacting nonlinear channels allow complex dynamical functions in open systems. For example, the competition between sodium and potassium channels with opposite gradients of ionic concentrations constructs a biological building block[1] for neural dynamics, including the Hopf bifurcation[14] with the 'all-or-none' characteristic and the repetitive firing[3] that enables the synchronization of neural networks[15]. Nonlinear resistance of memristors[2] has also been intensively applied to the electrical realization of chaotic systems[16] for encryption and random number generation. In these dynamical phenomena, the state-dependent nonlinear operation of each channel (*e.g.* potential-dependent gate dynamics[1]) and the interaction between channels (*e.g.* Kirchhoff's law for entire ionic currents[1]) determine the detailed form of the resulting dynamical function (*e.g.* threshold and repetitive firing in Hodgkin-Huxley neurons[1]).

Nonlinear channels for electromagnetic waves also lead to critical phenomena in nonequilibrium regimes, especially from saturable dynamics of amplification or absorption. In classical laser systems, the nonlinear response of the gain saturation is one of the key parameters determining the laser throughput[17], and the saturable absorber embedded in a laser cavity allows the mode locking and *Q*-switching for pulsed emission[7,8]. Recently, inspired by the concept of parity-time (PT) symmetry[18] in non-Hermitian quantum mechanics[19], the investigation on saturated optical channels has been revisited in terms of PT-symmetric phase transition[20,21]. The alteration of the PT-symmetric phase from the gain saturation has been explored to derive the interesting phenomena of non-reciprocal transparency of light[9,10] and robust wireless power transfer in radio-frequency circuits[11]. The evolution around the PT-symmetric exceptional point, also known as dynamical encircling, is now receiving growing attention for non-reciprocal mode conversion[13,22] with its topological robustness to nonlinear channels[12]. Notably, in contrast to competing ionic channels in Hodgkin-Huxley neurons[1], only a single nonlinear channel (saturable gain or loss) has been considered in the field of PT symmetry, missing possible important forms of dynamical functions from the interaction between channels.

In this paper, we investigate the competition between saturable gain and loss channels in PT-symmetric systems, and explore the dynamical phase diagram defined by equilibrium and oscillation parameters. We reveal that three different phases exist: fast- and slow-response equilibrium phases, and a nonequilibrium phase with weak oscillations. At the phase boundary between fast-response equilibrium and oscillatory nonequilibrium phases, the existence of the non-reciprocal strong oscillation state is demonstrated, which is a missing dynamical function in single nonlinear channel systems[9-13,22]. Utilizing the notion of the limit cycle in the real-valued parameter space[3,14], the chaotic nature of this phase boundary is analyzed for the applications of the intensity-based bifurcation and fully tunable time delay of repetitive resonator excitations. Due to their neuromorphic behaviors of threshold and repetitive firing, merged with PT-symmetric chaotic properties, the proposed PT-symmetric systems with competing saturable channels may become a building block for light-based artificial neural networks and encrypting systems.

The schematic of the dynamical PT-symmetric system is illustrated in Fig. 1a. The gain and loss resonators with field amplitudes $\psi_{G,L}$ are coupled each other with the coupling coefficient $\kappa$, while gain and loss channels possess intensity-dependent nonlinear saturations each with the gain[4] and loss[5] coefficients of $\gamma_{G,L} = \gamma_{G,L}(|\psi_{G,L}|) = \gamma_{G0,L0}/(1+|\psi_{G,L}/\psi_{Gs,Ls}|^2)$, where $\gamma_{G0,L0}$ are the linear coefficients, and $\psi_{Gs,Ls}$ are the saturation intensities. While the gain saturation generally occurs in pumped gain media (*e.g.* erbium-doped film[4,9]), the saturated loss can be obtained by utilizing organic dyes[23], carbon nanotube[24], graphene[25], and artificial realizations[26]. The amplification and absorption inside the resonators then decrease with increased field intensity (Fig. 1b), dynamically affecting the channels of the PT-symmetric system. From the temporal coupled mode theory[6], the governing equation of the PT-symmetric system is then written as

$$\frac{d}{dt}\begin{bmatrix}\psi_G(t)\\\psi_L(t)\end{bmatrix} = \begin{bmatrix} i\omega_0 + \gamma_{G0}\cdot\dfrac{1}{1+\left|\dfrac{\psi_G(t)}{\psi_{Gs}}\right|^2} & i\kappa \\ i\kappa & i\omega_0 - \gamma_{L0}\cdot\dfrac{1}{1+\left|\dfrac{\psi_L(t)}{\psi_{Ls}}\right|^2} \end{bmatrix}\begin{bmatrix}\psi_G(t)\\\psi_L(t)\end{bmatrix}, \qquad (1)$$

where $\omega_0$ is the resonant frequency of each isolated resonator.

The nonlinear temporal Eq. (1) can be solved numerically by applying the 6$^{th}$-order Runge-Kutta method[27] (time step = $2\pi/(200 \cdot \omega_0)$) with the initial condition of $\psi_{G,L}(t=0)$. To analyze dynamical behaviors of the system from the solutions $\psi_{G,L}(t)$, we define the characteristic parameters $\alpha$ and $\beta$, each measuring the equilibrium and oscillation condition of the system during the time range $T_1 \leq t \leq T_2$. In detail, for the field intensity of the entire system $I(t) = |\psi_G(t)|^2 + |\psi_L(t)|^2$, the 1$^{st}$ order polynomial fitting of $log[I(t)] \sim \alpha \cdot t + \delta$ presents the averaged equilibrium condition during $T_1 \leq t \leq T_2$: $\alpha = 0$ for the equilibrium phase, $\alpha > 0$ for amplifying, and $\alpha < 0$ for attenuating nonequilibrium phases with exponential evolutions. As well, inspired by the activity oscillation amplitude in neural networks[28], the temporal intensity oscillation inside the PT-symmetric system during $T_1 \leq t \leq T_2$ is then quantified by $\beta$, which measures the temporal deviation from the average signal activity $e^{\alpha \cdot t + \delta}$ as

$$\beta = \left[ \frac{1}{T_2 - T_1} \cdot \int_{T_1}^{T_2} \frac{|I(t) - e^{\alpha \cdot t + \delta}|^2}{|e^{\alpha \cdot t + \delta}|^2} dt \right]^{1/2}. \qquad (2)$$

where a high value of $\beta$ implies a high amplitude of the oscillations of the average activity inside the system[28].

Figures 1c and 1d each represent the calculated dynamical phase diagram of $\alpha$ and $\beta$, as functions of the relative strength ($\gamma_{L0}/\gamma_{G0}$) and saturation level ($\psi_{Ls}/\psi_{Gs}$) of the channels, with the initial condition of $\psi_G(t=0) = 1$ and $\psi_L(t=0) = 0$. From the dynamical characteristic parameters $\alpha$ and $\beta$, three distinct phases of nonequilibrium (N), fast-response- (E$_f$) and slow-response- (E$_s$) equilibriums are then observed with different equilibrium and oscillation conditions, also accompanying phase boundaries and a triple point. When the loss channel approaches the saturation level faster than the gain channel ($\psi_{Ls} < \psi_{Gs}$), the nonequilibrium phase N with the field amplification is obtained (Fig. 1e), while faster saturation of the gain channel ($\psi_{Gs} < \psi_{Ls}$) leads to the equilibrium phases E$_f$ and E$_s$. The equilibrium phases can then be divided into two distinct phases of E$_f$ with fast temporal response (Fig. 1f) and E$_s$ with slow temporal response (Fig. 1g) according to the relative strength of the gain and loss channels ($\gamma_{L0}/\gamma_{G0}$). It is also noted that a single nonlinear channel system[9-13] ($\psi_{Ls} \to \infty$) corresponds to a single 'line' in those two-dimensional phase diagrams.

In contrast to continuous phase transition in linear PT-symmetric systems[20,21], the transition between different dynamical phases occurs discontinuously across phase boundaries. For example, considering that the quasi-static phase of PT symmetry (see Supplementary Note S1) is unbroken in N and $E_s$ regimes, and is broken in the $E_f$ regime, exotic 'phase boundaries' in terms of the oscillation are observed between unbroken and broken PT-symmetric phases of N-$E_f$ or $E_f$-$E_s$ (red filled symbols in Fig. 1c,d). As shown in those boundaries in Fig. 1d, strong dynamical oscillation (with large $\beta$) inside the system can be achieved at the boundary between unbroken and broken PT-symmetric phases, having distinct eigenmodal profiles and real-complex eigenvalues[20].

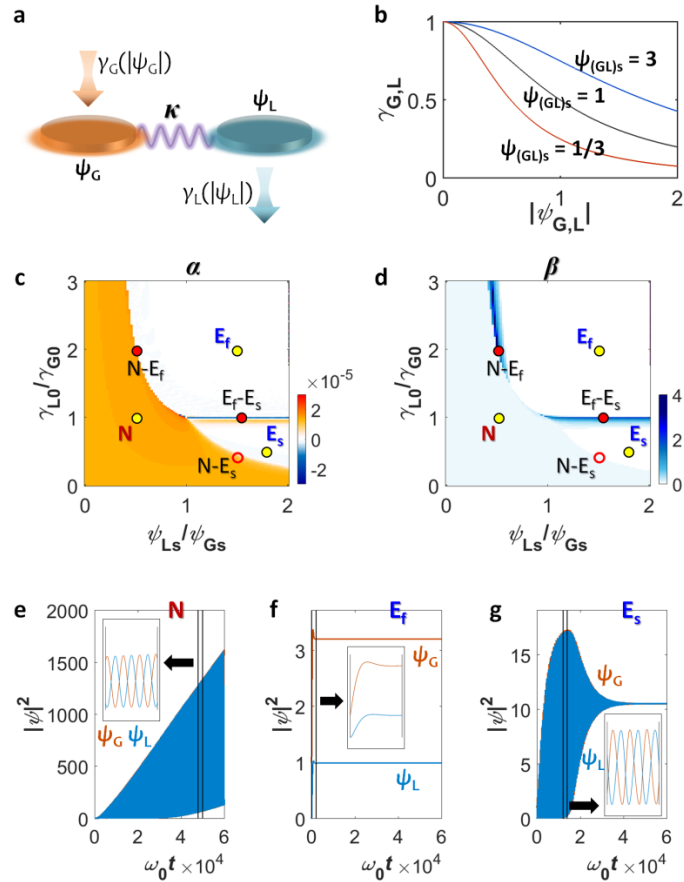

**Figure 1. Dynamical phase diagram of the PT-symmetric system with competing saturable channels.** (a) A schematic for the system with saturable channels, from the nonlinear coefficient functions $\gamma_{G,L} = \gamma_{G,L}(|\psi_{G,L}|)$. (b)

The field-amplitude-dependent variation of the gain and loss coefficients, for different saturation levels ($\psi_{Gs}, \psi_{Ls}$). Dynamical phase diagrams defined by (c) the equilibrium parameter $\alpha$ and (d) the oscillation parameter $\beta$: N for the nonequilibrium phase, and $E_f$ (or $E_s$) for the equilibrium phase of the fast (or slow) temporal response. Yellow filled circles in (c,d) present the cases in (e-g), for the temporally varying field intensities inside the gain ($\psi_G$) and loss ($\psi_L$) resonators. Red filled circles in (c,d) present the case of the oscillating phase boundaries N-$E_f$ and $E_f$-$E_s$, while a red empty circle denotes the non-oscillating phase boundary N-$E_s$. The initial condition is $\psi_G(t=0) = 1$ and $\psi_L(t=0) = 0$ for (c-g). The time range for calculating $\alpha$ and $\beta$ is $[T_1, T_2] = [1\times10^4, 5\times10^4]\cdot\pi/\omega_0$. $\kappa = \gamma_{G0} = 5\times10^{-3}$, and $\psi_{Gs} = 2$ for all cases.

Among two phase boundaries N-$E_f$ and $E_f$-$E_s$ with strong oscillations (or large $\beta$), we now focus on the phase boundary N-$E_f$, which offers the exotic and unique dynamical function of non-reciprocal strong oscillations, absent in the phase boundary $E_f$-$E_s$, the triple point N-$E_f$-$E_s$, and the single channel system[9-13,22] (see Supplementary Note S2-S4). We first analyze the behavior of the phase boundary N-$E_f$, in terms of the initial excitation channel. Figure 2a-d represents the enlarged $\alpha$-$\beta$ phase diagrams around the phase boundary N-$E_f$ for the initial excitation conditions of $\psi_G(t=0) = 1$, $\psi_L(t=0) = 0$ (Fig. 2a,b) and $\psi_G(t=0) = 0$, $\psi_L(t=0) = 1$ (Fig. 2c,d). Due to the discontinuity at the boundary N-$E_f$, the switching of 'phase' with strong dependence on the initial excitation condition is observed in Fig. 2a-d, for both $\alpha$ (from N to $E_f$ phase) and $\beta$ (from non-oscillating to oscillating state) parameters (red dashed lines). In detail, for the gain-resonator excitation (Fig. 2a,b), the overall system remains at the N phase with a low value of $\beta$, leading to the weakly-oscillating amplifying response (Fig. 2e). In contrast, for the loss-resonator excitation (Fig. 2c,d), the system stays at the equilibrium phase $E_f$ with a high value of $\beta$. This large $\beta$ with $\alpha = 0$ corresponds to the repetitive excitation of both resonators (Fig. 2f) with the oscillation across the exceptional point (see Supplementary Note S5 for oscillating PT-symmetric transition).

This 'chaotic' phenomenon with a highly sensitive response to the initial excitation condition originates from the non-reciprocity of the dynamical PT-symmetric system, furthermore, resembling a repetitive firing response of Hodgkin-Huxley neurons[1,3]: the analogy of sodium (incoming current) and potassium (outgoing current) gated channels, with gain (amplified light) and loss (dissipative light) saturated channels.

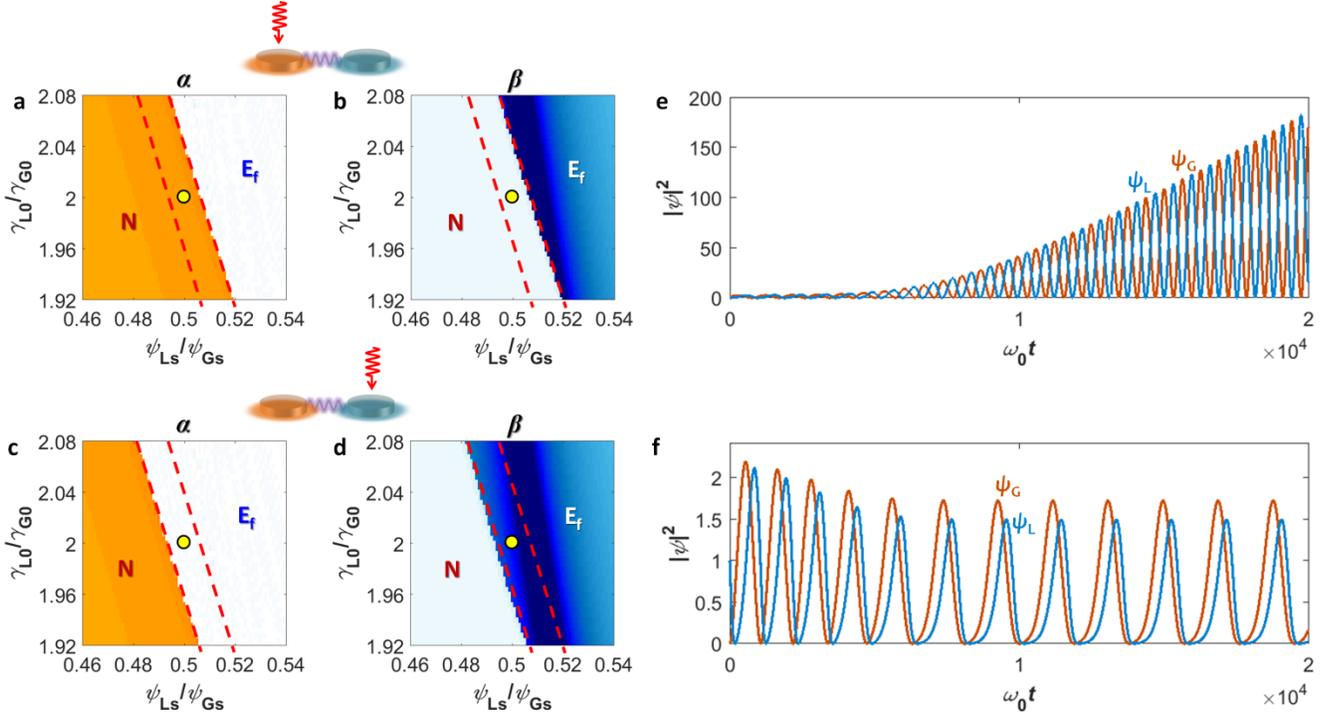

**Figure 2. Non-reciprocal emergence of repetitive resonator firing of the PT-symmetric system.** The enlarged phase diagrams around the phase boundary N-$E_f$ in the cases of (a,b) gain-resonator excitation and (c,d) loss-resonator excitation, for (a,c) the equilibrium parameter $\alpha$ and (b,d) the oscillation parameter $\beta$. (e,f) The temporally varying field intensities in the gain ($\psi_G$) and loss ($\psi_L$) resonators for each case of the yellow filled circles in Fig. 2a,b and Fig. 2c,d. All other parameters are the same as those in Fig. 1.

To further look into the chaotic, oscillatory behavior at the phase boundary N-$E_f$, we now develop the limit cycle analysis which has been utilized for analyzing repetitive firing in Hodgkin-Huxley neurons[3,14]. For the field amplitude $\psi_{G,L}$, we employ the representation based on real-valued parameters[29] $I(t)$, $\theta(t)$, $\phi(t)$, and $\omega(t)$, as $\psi_G(t) = [I(t)]^{1/2} \cdot cos[\theta(t)] \cdot exp[+i\phi(t)/2] \cdot exp[i\omega(t)]$ and $\psi_L(t) = [I(t)]^{1/2} \cdot sin[\theta(t)] \cdot exp[-i\phi(t)/2] \cdot exp[i\omega(t)]$. With some elaborations (see Supplementary Note S6), Eq. (1) becomes the closed equation of the form of

$$\begin{aligned}
\frac{dI}{dt} &= I \cdot \sin 2\theta \cdot \left( \gamma_{G0} \cdot \frac{I_{Gs} \cdot \cot \theta}{I_{Gs} + I \cos^2 \theta} - \gamma_{L0} \cdot \frac{I_{Ls} \cdot \tan \theta}{I_{Ls} + I \sin^2 \theta} \right) \\
\frac{d\theta}{dt} &= -\kappa \sin \phi - \frac{\sin 2\theta}{2} \cdot \left( \gamma_{G0} \cdot \frac{I_{Gs}}{I_{Gs} + I \cos^2 \theta} + \gamma_{L0} \cdot \frac{I_{Ls}}{I_{Ls} + I \sin^2 \theta} \right), \\
\frac{d\phi}{dt} &= -2\kappa \cdot \cot 2\theta \cdot \cos \phi
\end{aligned} \quad (3)$$

with the additional detuning equation $d\omega/dt = \omega_0 + \kappa \cdot cos(\phi)/sin(2\theta)$, where $I_{Gs} = |\psi_{Gs}|^2$ and $I_{Ls} = |\psi_{Ls}|^2$.

The evolutions of optical states are shown in Fig. 3a,b, each for the gain and loss resonator excitation. It is noted that the directional rotation in the $\theta$-$\phi$ parameter space occurs for both cases, which corresponds to the alternating transition between amplitude ($\theta$) and phase ($\varphi$) differences of the field in each resonator, in accordance with the oscillating PT-symmetric phase transitions (Supplementary Note S5). However, depending on the initial excitation channel (red and blue symbols in the $I(t=0) = 1$, $\theta$-$\phi$ green plane of Fig. 3), the bifurcation of the system between the nonequilibrium state (Fig. 3a) and the equilibrium state with the limit cycle operation (Fig. 3b) is evident, representing the 'open' and 'closed' cycles for each case in the $\theta$-$I$ plane (bottom figures in Fig. 3). This non-reciprocity originates from the distinct initial variation of gain-loss distributions determined by the excited channel, leading to the chaotic state between weakly-oscillating nonequilibrium and strongly-oscillating equilibrium phases (Fig. 2e,3a versus 2f,3b).

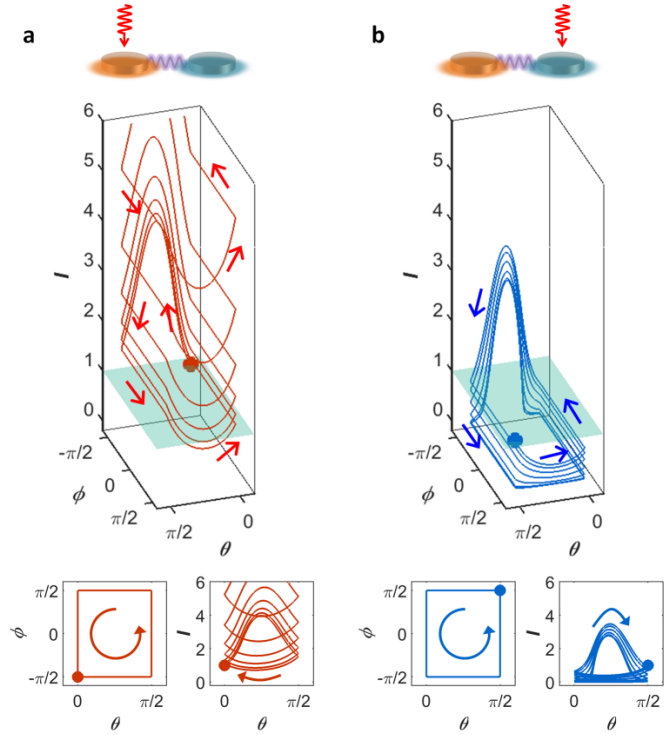

**Figure 3. Non-reciprocity of the dynamical evolutions of optical states in the real-valued parameter space.** Temporal variation of optical states in the $I$-$\theta$-$\phi$ space for the (a) gain-resonator excitation and (b) loss-resonator

excitation. Bottom figures present the trajectories on the $\theta$-$\phi$ and $\theta$-$I$ planes. Solid circles denote the initial point and arrows present the direction of the evolution.

The chaotic feature on the initial excitation channel can be extended to the intensity-dependent initial condition with different $I(t=0)$. Figure 4a-c represents the intensity-dependent dynamical evolution, when the loss resonator is excited. Although the initial conditions of $I(t=0) \leq 1$ eventually lead to the convergence to the final locus of the limit cycle, the 'speed' of the convergence depends on the initial field intensity, deriving slower convergence for larger intensity. With the identical limit cycles that have different convergence time, the delicate control of the time delay of the repetitive resonator firing (Fig. 4d) becomes possible, where the form and magnitude of the field excitation are still preserved. As shown in Fig. 4d, the control of the initial field intensity allows the full coverage of the 'temporal phase' of the repetitive resonator firing.

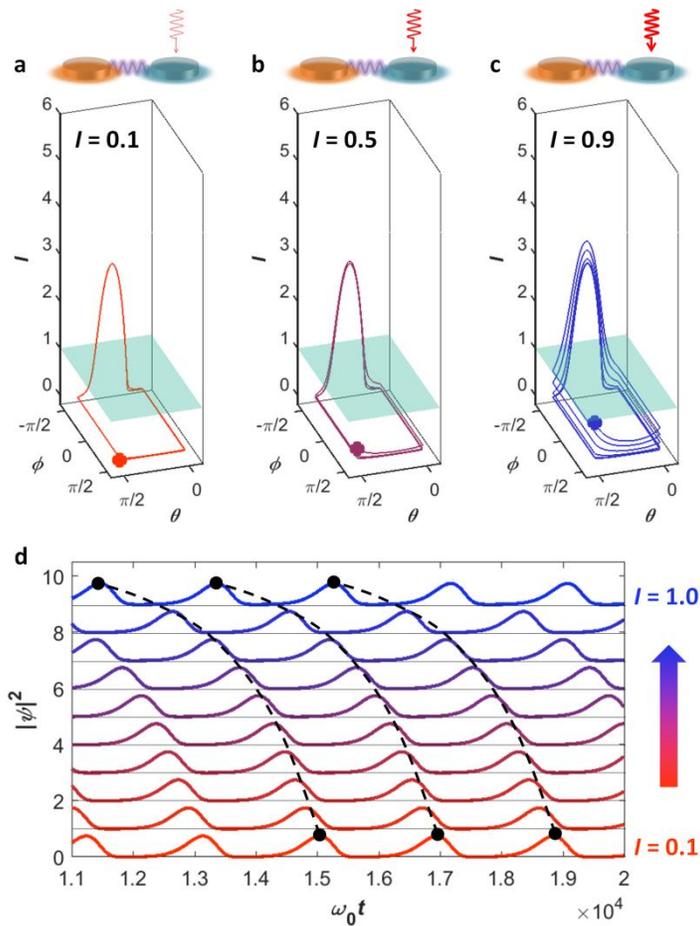

**Figure 4. Intensity-based control of the time delay for repetitive resonator firing.** Dynamical evolutions of optical states for (a) $I(t=0) = 0.1$, (b) $I(t=0) = 0.5$, and (c) $I(t=0) = 0.9$. Solid circles denote the initial state of each case. (d) The temporally varying field intensities $I(t) = |\psi_G(t)|^2 + |\psi_L(t)|^2$ for the initial condition $I(t=0) = 0.1$ to 1.0. Black dashed lines show the evolution of the excitation peak, controlled by the excited field intensity. Only the loss resonator is excited at $t = 0$. All other parameters are the same as those in Fig. 1.

In summary, we have shown that the competing phenomenon between saturable gain and loss channels is described in the form of the dynamical phase diagram. Each phase exhibits different equilibrium and oscillation features, in relation to the quasi-static variation of PT-symmetric gauges and phases. At the abrupt phase boundary between these dynamical phases, the chaotic state with non-reciprocity exists with the realization of non-reciprocal repetitive firing in resonators. From the chaotic feature, the convergence speed to the limit cycle can also be manipulated, allowing the fully tunable time delay of the resonator excitation. Considering the practical realization from the controlled saturation with engineered energy bands[30] or artificial realizations[26], the fully tunable repetitive firing of electromagnetic waves may become a proper building block for neuromorphic wave circuits, underpinning the critical role of the competition between nonlinear dynamics for high-level functions or network synchronizations.

**Acknowledgments**

This work was supported by the National Research Foundation of Korea (NRF) through the Global Frontier Program (GFP, 2014M3A6B3063708) funded by the Korean government. S. Yu was supported by the Basic Science Research Program (2016R1A6A3A04009723), and X. Piao and N. Park were supported by the Korea Research Fellowship Program (KRF, 2016H1D3A1938069) through the NRF, all funded by the Korean government.

**Author Contributions**

S.Y. conceived the presented idea. S.Y. and X.P. developed the theory and performed the computations. N.P. encouraged S.Y. to investigate the optical analogy of neural systems while supervising the findings of this work. All authors discussed the results and contributed to the final manuscript.

**Competing Interests Statement**

The authors declare that they have no competing financial interests.

# References

1. Hodgkin, A. L. & Huxley, A. F. A quantitative description of membrane current and its application to conduction and excitation in nerve. *J. Physiol.* **117**, 500 (1952).

2. Strukov, D. B., Snider, G. S., Stewart, D. R. & Williams, R. S. The missing memristor found. *Nature* **453**, 80-83 (2008).

3. Guttman, R., Lewis, S. & Rinzel, J. Control of repetitive firing in squid axon membrane as a model for a neuroneoscillator. *J. Physiol.* **305**, 377 (1980).

4. Yang, L., Carmon, T., Min, B., Spillane, S. M. & Vahala, K. J. Erbium-doped and Raman microlasers on a silicon chip fabricated by the sol–gel process. *Appl. Phys. Lett.* **86**, 091114 (2005).

5. Hercher, M. An analysis of saturable absorbers. *Appl. Opt.* **6**, 947-954 (1967).

6. Haus, H. A. *Waves and fields in optoelectronics*. (Prentice-Hall Englewood Cliffs, NJ, 1984).

7. Haus, H. A. Theory of mode locking with a fast saturable absorber. *J. Appl. Phys.* **46**, 3049-3058 (1975).

8. Haus, H. Theory of mode locking with a slow saturable absorber. *IEEE J. Quantum Electron.* **11**, 736-746 (1975).

9. Chang, L., Jiang, X., Hua, S., Yang, C., Wen, J., Jiang, L., Li, G., Wang, G. & Xiao, M. Parity–time symmetry and variable optical isolation in active–passive-coupled microresonators. *Nat. Photon.* **8**, 524-529 (2014).

10. Peng, B., Özdemir, Ş. K., Lei, F., Monifi, F., Gianfreda, M., Long, G. L., Fan, S., Nori, F., Bender, C. M. & Yang, L. Parity–time-symmetric whispering-gallery microcavities. *Nat. Phys.* **10**, 394-398 (2014).

11. Assawaworrarit, S., Yu, X. & Fan, S. Robust wireless power transfer using a nonlinear parity–time-symmetric circuit. *Nature* **546**, 387-390 (2017).

12. Hassan, A. U., Zhen, B., Soljačić, M., Khajavikhan, M. & Christodoulides, D. N. Dynamically Encircling Exceptional Points: Exact Evolution and Polarization State Conversion. *Phys. Rev. Lett.* **118**, 093002 (2017).

13. Choi, Y., Hahn, C., Yoon, J. W., Song, S. H. & Berini, P. Extremely broadband, on-chip optical nonreciprocity enabled by mimicking nonlinear anti-adiabatic quantum jumps near exceptional points. *Nat. Commun.* **8**, 14154 (2017).

14. Wang, J., Chen, L. & Fei, X. Bifurcation control of the Hodgkin–Huxley equations. *Chaos, Solitons & Fractals* **33**, 217-224 (2007).

15. Wang, W., Chen, G. & Wang, Z. 40-Hz coherent oscillations in neuronal systems. *Phys. Rev. E* **56**, 3728 (1997).

16. Muthuswamy, B. & Kokate, P. P. Memristor-based chaotic circuits. *IETE Tech. Rev.* **26**, 417-429 (2009).

17. Yariv, A. & Yeh, P. *Photonics: optical electronics in modern communications*. (Oxford University Press 2007).

18. Bender, C. M. & Boettcher, S. Real spectra in non-Hermitian Hamiltonians having PT symmetry. *Phys. Rev. Lett.* **80**, 5243 (1998).

19. Bender, C. M., Brody, D. C. & Jones, H. F. Complex Extension of Quantum Mechanics. *Phys. Rev. Lett.* **89**, 270401 (2002).

20. Rüter, C. E., Makris, K. G., El-Ganainy, R., Christodoulides, D. N., Segev, M. & Kip, D. Observation of parity–time symmetry in optics. *Nat. Phys.* **6**, 192-195 (2010).

21. Guo, A., Salamo, G., Duchesne, D., Morandotti, R., Volatier-Ravat, M., Aimez, V., Siviloglou, G. & Christodoulides, D. Observation of PT-symmetry breaking in complex optical potentials. *Phys. Rev. Lett.* **103**, 093902 (2009).

22. Doppler, J., Mailybaev, A. A., Böhm, J., Kuhl, U., Girschik, A., Libisch, F., Milburn, T. J., Rabl, P., Moiseyev, N. & Rotter, S. Dynamically encircling an exceptional point for asymmetric mode switching. *Nature* **537**, 76-79 (2016).

23. Spiller, E. Saturable optical resonator. *J. Appl. Phys.* **43**, 1673-1681 (1972).

24. Set, S. Y., Yaguchi, H., Tanaka, Y. & Jablonski, M. Laser mode locking using a saturable absorber incorporating carbon nanotubes. *J. Lightwave Technol.* **22**, 51 (2004).

25. Bao, Q., Zhang, H., Wang, Y., Ni, Z., Yan, Y., Shen, Z. X., Loh, K. P. & Tang, D. Y. Atomic-layer graphene as a saturable absorber for ultrafast pulsed lasers. *Adv. Funct. Mater.* **19**, 3077-3083 (2009).

26. Teimourpour, M., Rahman, A., Srinivasan, K. & El-Ganainy, R. Non-Hermitian Engineering of Synthetic Saturable Absorbers for Applications in Photonics. *Phys. Rev. Applied.* **7**, 014015 (2017).

27. Butcher, J. C. On Runge-Kutta processes of high order. *J. Austral. Math. Soc.* **4**, 179-194 (1964).

28. Lago-Fernández, L. F., Huerta, R., Corbacho, F. & Sigüenza, J. A. Fast response and temporal coherent oscillations in small-world networks. *Phys. Rev. Lett.* **84**, 2758 (2000).

29. Sukhorukov, A. A., Xu, Z. & Kivshar, Y. S. Nonlinear suppression of time reversals in PT-symmetric optical couplers. *Phys. Rev. A* **82**, 043818 (2010).

30. Rafailov, E. U., Cataluna, M. A. & Avrutin, E. A. *Ultrafast lasers based on quantum dot structures: physics and devices*. (John Wiley & Sons, 2011).

# Supplementary Information for "Dynamical phase diagram of parity-time symmetry with competing saturable channels"

Sunkyu Yu, Xianji Piao, and Namkyoo Park*

*Photonic Systems Laboratory, Department of Electrical and Computer Engineering, Seoul National University, Seoul 08826, Korea*

*E-mail address for correspondence: nkpark@snu.ac.kr

**Note S1. Quasi-static analysis of N, $E_f$, and $E_s$ phases**

**Note S2. Quasi-reciprocal phase boundary $E_f$-$E_s$**

**Note S3. Triple point N-$E_f$-$E_s$**

**Note S4. Comparison with a single-nonlinear-channel system**

**Note S5. Quasi-static analysis of the phase boundary N-$E_f$**

**Note S6. Derivation of the limit cycle equation**

## Note S1. Quasi-static analysis of N, $E_f$, and $E_s$ phases

For the comprehensible understanding of the dynamical PT-symmetric system, in this note, we assume the adiabatic evolution of the system, which enables the quasi-static analysis. For Eq. (1) in the main text, the instantaneous eigenvalues $\omega_e(t)$ of the system then becomes

$$\omega_e(t) = \omega_0 - i \cdot \frac{\dfrac{\gamma_{G0}}{1+\left|\dfrac{\psi_G(t)}{\psi_{Gs}}\right|^2} - \dfrac{\gamma_{L0}}{1+\left|\dfrac{\psi_L(t)}{\psi_{Ls}}\right|^2}}{2} \pm \left[\kappa^2 - \left(\dfrac{\dfrac{\gamma_{G0}}{1+\left|\dfrac{\psi_G(t)}{\psi_{Gs}}\right|^2} + \dfrac{\gamma_{L0}}{1+\left|\dfrac{\psi_L(t)}{\psi_{Ls}}\right|^2}}{2}\right)^2\right]^{1/2}, \quad (S1)$$

which can be simplified to $\omega_e(t) = \omega_0 - i \cdot \kappa \cdot \gamma_{avg}(t) \pm \kappa \cdot D(t)^{1/2}$ where

$$\gamma_{avg}(t) = \frac{\dfrac{\gamma_{G0}}{1+\left|\dfrac{\psi_G(t)}{\psi_{Gs}}\right|^2} - \dfrac{\gamma_{L0}}{1+\left|\dfrac{\psi_L(t)}{\psi_{Ls}}\right|^2}}{2\kappa},$$

$$D(t) = 1 - \left(\dfrac{\dfrac{\gamma_{G0}}{1+\left|\dfrac{\psi_G(t)}{\psi_{Gs}}\right|^2} + \dfrac{\gamma_{L0}}{1+\left|\dfrac{\psi_L(t)}{\psi_{Ls}}\right|^2}}{2\kappa}\right)^2. \quad (S2)$$

The temporal function $\gamma_{avg}(t)$ determines the gauge of the PT-symmetric system[1,2], as $\gamma_{avg} > 0$ for the active gauge, $\gamma_{avg} < 0$ for the passive gauge, and $\gamma_{avg} = 0$ for the exact PT symmetry. Meanwhile, $D(t)$ defines the phase of PT symmetry[3]: $D(t) > 0$ for unbroken phase, $D(t) < 0$ for broken phase, and $D(t) = 0$ for exceptional point.

Figure S1 shows the temporal evolution of gauges $\gamma_{avg}(t)$ (Fig. S1a-c) and PT-symmetric phases $D(t)$ (Fig. S1d-f) in the regimes of N, $E_f$, and $E_s$ phases of the dynamical PT-symmetric system, with the initial condition of $\psi_G(t=0) = 1$ and $\psi_L(t=0) = 0$. Firstly, N and $E_s$ phases which have the strong oscillations of $\gamma_{avg}(t)$ and $D(t)$ maintain the unbroken PT-symmetric phase (Fig. S1d,f), due to the power exchange from two distinct eigenvalues with different eigenmodes in the unbroken PT symmetry[1,3]. In contrast, the phase of $E_f$ corresponds to the broken PT-symmetric phase with the identical real part of eigenvalues[1,3] (Fig. S1e), leading to the rapid convergence to the steady state. From the gauge analysis, it is also shown that the nonequilibrium property of the N phase originates from the spiking of the gauge $\gamma_{avg}(t)$ of the system (enlarged plot of Fig. S1a). However, as shown in the similarity of the gauge and PT-symmetric phase between N and $E_s$ phases (Fig. S1a,d versus Fig. S1c,f), it is noted that the dynamical phase diagrams of $\alpha$ and $\beta$ provide deeper understanding of the temporal dynamics with the clear distinction of the nonequilibrium (N) and equilibrium ($E_s$) phases.

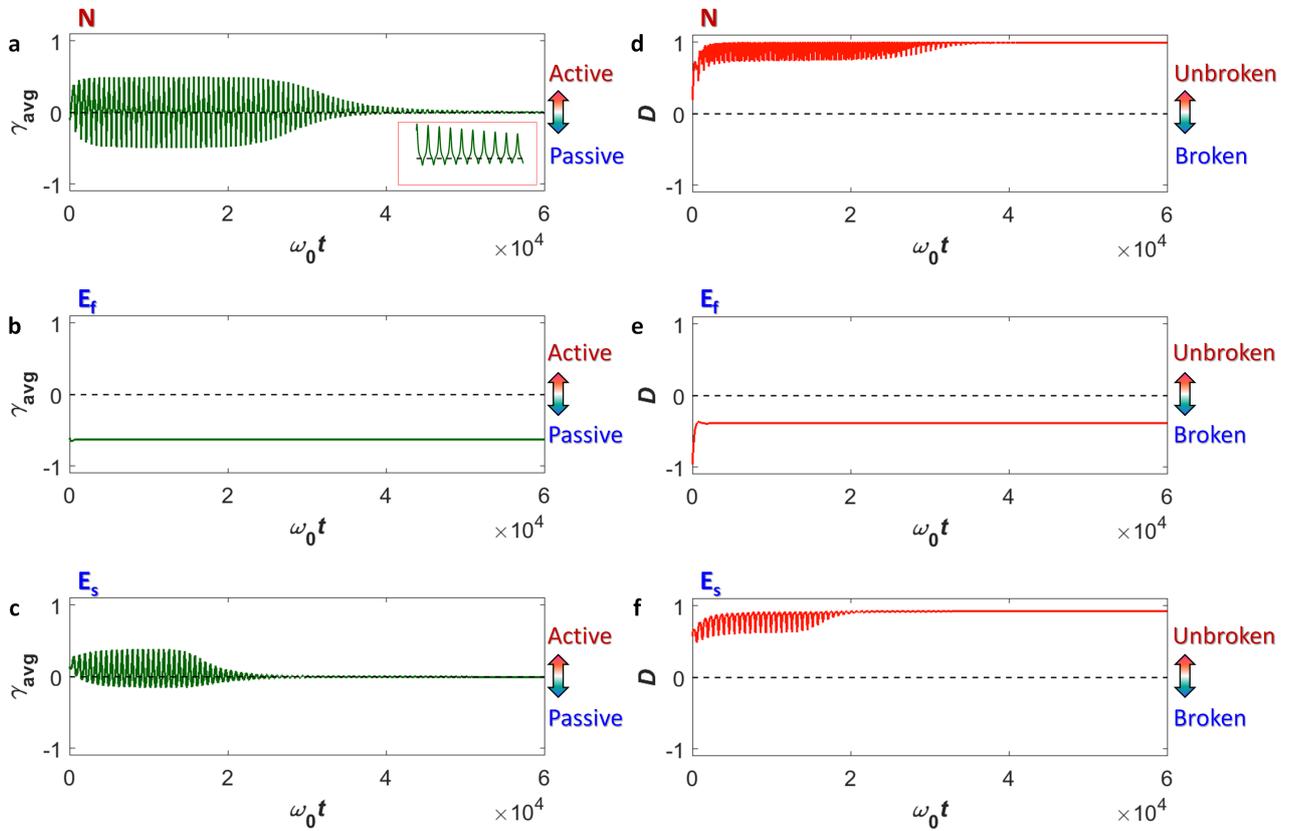

**Figure S1. Quasi-static analysis of N, $E_f$, and $E_s$ phases.** (a-c) Temporal gauges $\gamma_{avg}(t)$ and (d-f) PT symmetry phase parameter $D(t)$ for (a,d) N, (b,e) $E_f$, and (c,f) $E_s$ phases. The dynamical gain and loss coefficients are calculated from the 6$^{th}$-order Runge-Kutta analysis, for the calculation of Eq. (S2). All other parameters are the same as in Fig. 1 in the main text.

## Note S2. Quasi-reciprocal phase boundary $E_f$-$E_s$

The phase boundary N-$E_f$ in between the nonequilibrium (N) and equilibrium ($E_f$) phases is highly sensitive to the initial condition, leading to the chaotic property and the following non-reciprocity, intensity-based bifurcation and time delay (see Fig. 2-4 in the main text). In contrast, at the phase boundary $E_f$-$E_s$ in between equilibrium ($E_f$ and $E_s$) phases, the overall energy inside the system is converged, with the significantly suppressed chaotic nature.

Figure S2a-d represents the enlarged phase diagrams of $\alpha$ and $\beta$ around the phase boundary $E_f$-$E_s$, for the initial condition of $\psi_G(t=0) = 1$, $\psi_L(t=0) = 0$ (Fig. S2a,b) and $\psi_G(t=0) = 0$, $\psi_L(t=0) = 1$ (Fig. S2c,d). Because of the equilibrium condition in $E_f$ and $E_s$ phases, the initial condition does not affect much the nonlinear gain and loss channels, leading to quasi-reciprocal responses (Fig. S2a,b versus Fig. S2c,d). However, around the phase boundary $E_f$-$E_s$ which is near the exceptional point of linear PT-symmetric systems ($D(t) = 0$ when $\gamma_{G0} = \gamma_{L0}$, for $\kappa = \gamma_{G0}$ and $\psi_{Gs,Ls} \to \infty$, Eq. (S2)), the equilibrium condition of $\alpha$ varies drastically (Fig. S2a,c).

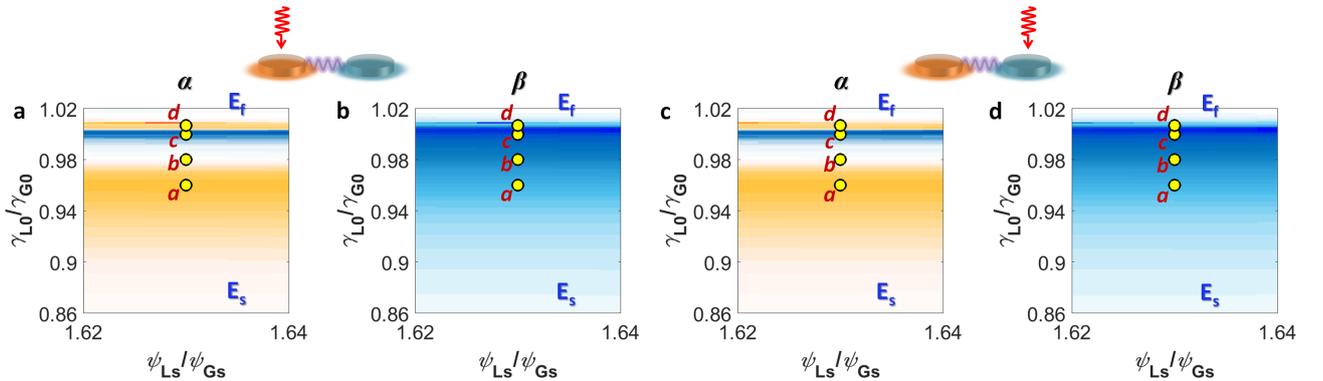

**Figure S2. Quasi-reciprocal response at the phase boundary $E_f$-$E_s$.** The enlarged phase diagrams around the phase boundary $E_f$-$E_s$ in the cases of (a,b) gain-resonator excitation and (c,d) loss-resonator excitation, for (a,c) the equilibrium parameter $\alpha$ and (b,d) the oscillation parameter $\beta$. Each yellow filled circle denotes the distinguished case in the equilibrium phase diagram $\alpha$, with the temporal evolutions in Fig. S3. All other parameters are the same as those in Fig. 1.

Figure S3 shows the temporal evolution at each point near the phase boundary $E_f$-$E_s$. While preserving quasi-reciprocal responses, the convergences to non-oscillating constant states (Fig. S3a,d) and oscillating states (Fig. S3b,c) are observed with relatively slower response than that at the phase boundary N-$E_f$ (Fig. 2e,f in the main text).

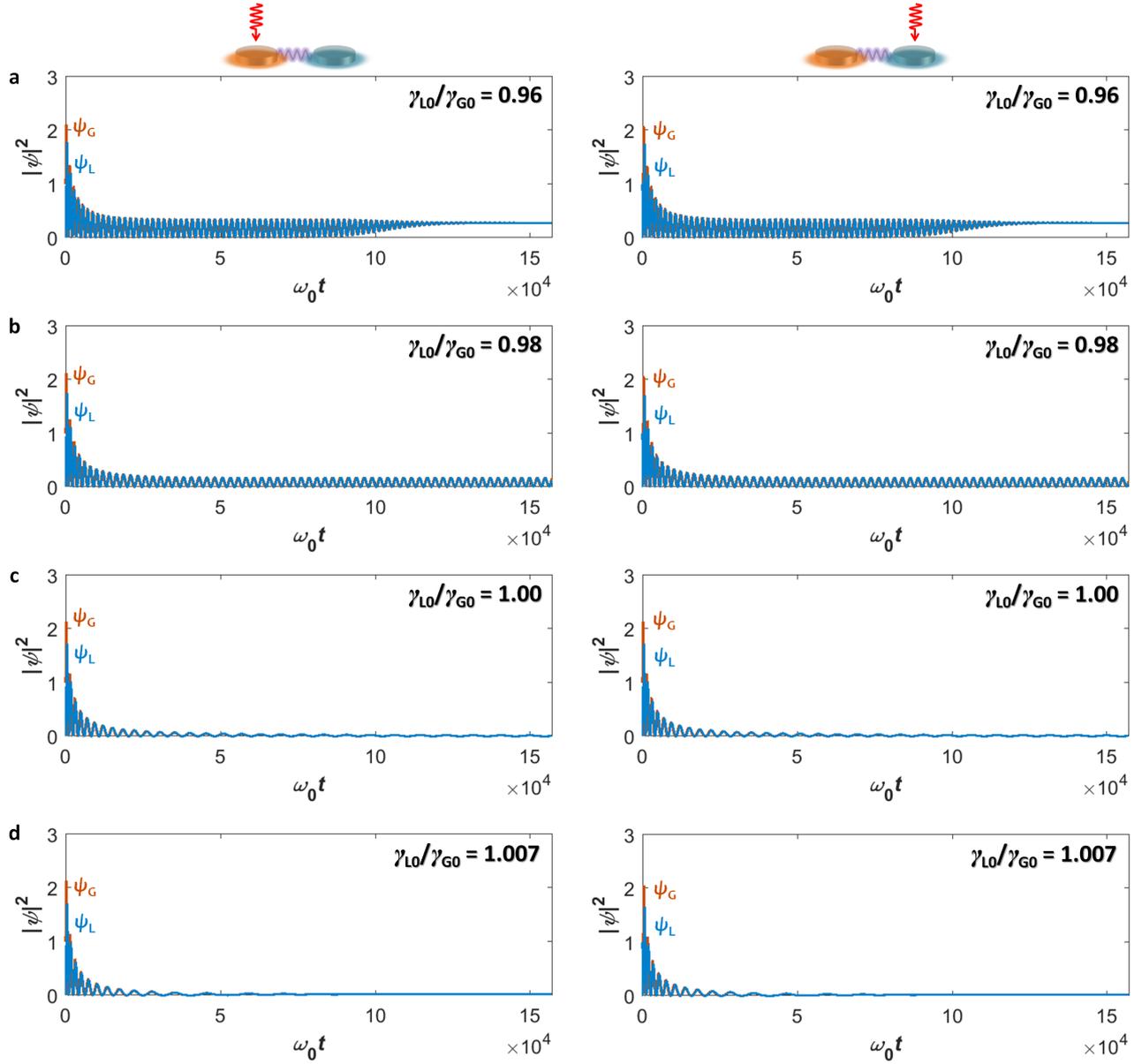

**Figure S3. Temporal dynamics near the phase boundary $E_f$-$E_s$.** (a-d) The temporally varying field intensities in the gain ($\psi_G$) and loss ($\psi_L$) resonators, for each case of the yellow filled circles in Fig. S2, with different values of the relative strength of the loss channel $\gamma_{L0}/\gamma_{G0}$: (left) gain-resonator excitation and (right) loss-resonator excitation. All other parameters are the same as those in Fig. 1.

## Note S3. Triple point N-$E_f$-$E_s$

Figure S4 shows the enlarged phase diagrams of $\alpha$ and $\beta$ near the triple point N-$E_f$-$E_s$. While the response is quasi-reciprocal (Fig. S4a,b versus Fig. S4c,d), the equilibrium phase diagram varies drastically between the equilibrium phase ($\alpha = 0$), amplifying ($\alpha > 0$), and attenuating ($\alpha < 0$) nonequilibrium phases, as the intrinsic property of the triple point in phase diagrams.

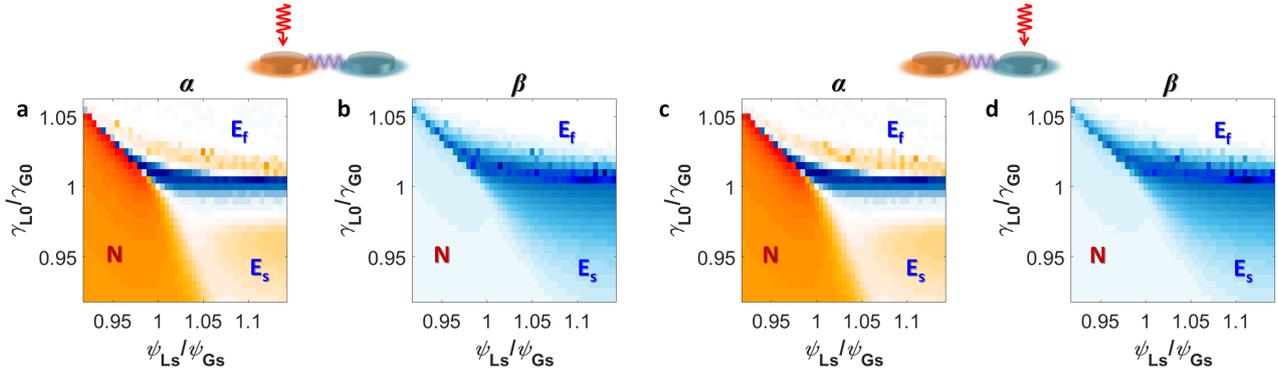

**Figure S4. Phase diagrams near the triple point.** The enlarged phase diagrams around the triple point N-$E_f$-$E_s$ in the cases of (a,b) gain-resonator excitation and (c,d) loss-resonator excitation, for (a,c) the equilibrium parameter $\alpha$ and (b,d) the oscillation parameter $\beta$. All other parameters are the same as in Fig. 1 in the main text.

## Note S4. Comparison with a single-nonlinear-channel system

In the two-dimensional phase diagrams from competing saturable channels (Fig. S5a), a single nonlinear channel system[4-8] corresponds to the single 'line'. We compare the regimes including the phase boundaries of N-$E_f$ (Fig. S5b) and $E_f$-$E_s$ (Fig. S5c) with the single channel system (Fig. S5d), which has the gain saturation and linear loss ($\psi_{Ls} \to \infty$, red dashed lines in Fig. S5a). In Fig. S5b-d, the uniqueness of the non-reciprocity (difference between red and blue lines) and strong oscillation (large $\beta$) is evident only at the phase boundary N-$E_f$, when compared with the quasi-reciprocal and weakly oscillating responses at the phase boundary $E_f$-$E_s$ and in the single channel case.

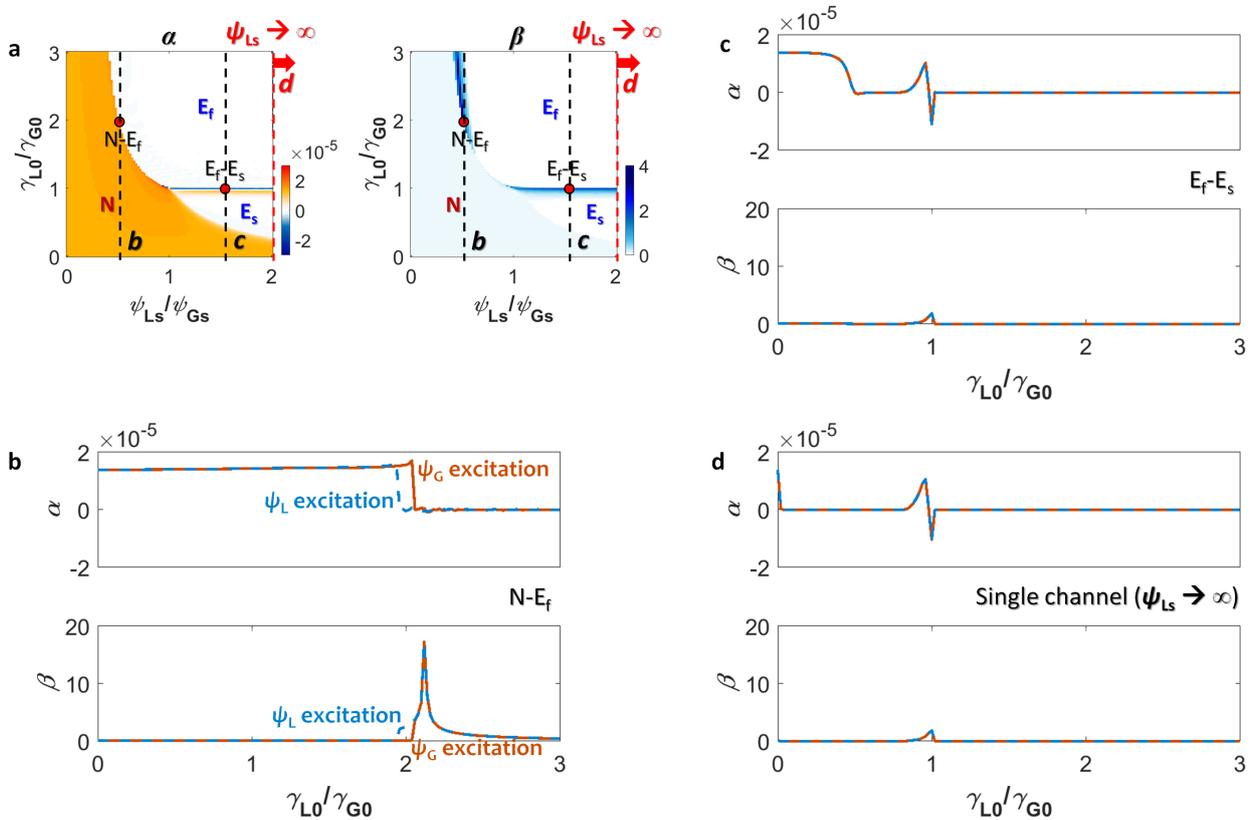

**Figure S5. Comparison with a single nonlinear channel system.** (a) Phase diagrams of $\alpha$ and $\beta$, representing the plotted regimes in (b-d). Black dashed lines denote the regimes including the phase boundaries of N-$E_f$ and $E_f$-$E_s$. Red dashed line denotes the single nonlinear channel system. (b-d) The evolutions of $\alpha$ and $\beta$, as a function of the relative strength ($\gamma_{L0}/\gamma_{G0}$) of the channels for the regimes across (b) N-$E_f$, (c) $E_f$-$E_s$, and (d) for the single channel case. Red lines represent the cases of the gain resonator excitation, and blue lines represent the cases of the loss resonator excitation. All other parameters are the same as in Fig. 1 in the main text.

## Note S5. Quasi-static analysis of the phase boundary N-$E_f$

The results of the quasi-static analysis for the phase boundary N-$E_f$ are shown in Fig. S6, in terms of gauges $\gamma_{avg}(t)$ (Fig. S6a,b) and PT-symmetric phases $D(t)$ (Fig. S6c,d). When the gain resonator is excited ($\psi_G(t=0) = 1$ and $\psi_L(t=0) = 0$, Fig. S6a,c), there exists the oscillation between active and passive gauges (Fig. S6a) which results in the nonequilibrium condition as shown in the main text. It is also noted that because the PT-symmetric phase is at the unbroken regime (Fig. S6c, $D(t) > 0$), the evolution of optical energy has the form of the directional coupling[3] (Fig. 2e in the main text).

In contrast, for the case of the loss-resonator excitation, while the passive gauge is maintained (Fig. S6b), the oscillation of PT-symmetric phase occurs around the exceptional point (Fig. S6d, $D(t) \sim 0$) which directly determines the form of the repetitive resonator excitation (Fig. 2f in the main text). For example, because of the evolution near the exceptional point, two resonators are almost synchronized from the coalescence of eigenmodes[3].

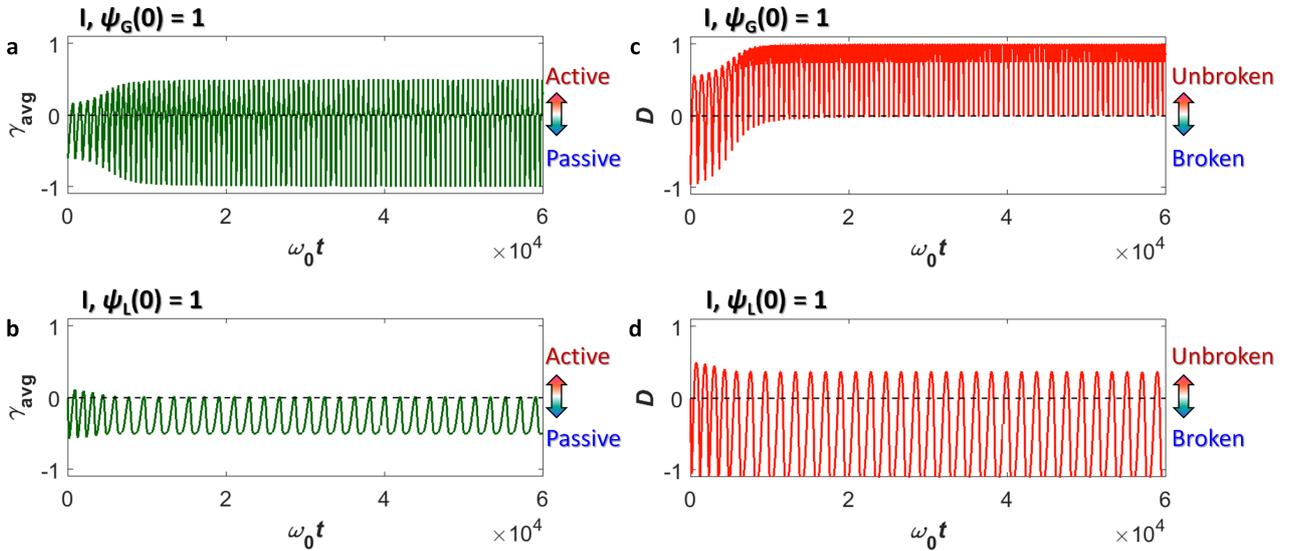

**Figure S6. Quasi-static analysis of the phase boundary N-$E_f$.** (a,b) Temporal gauges $\gamma_{avg}(t)$ and (c,d) PT symmetry phase parameter $D(t)$ for the (a,c) gain-resonator and (b,d) loss-resonator excitations. The saturated gain and loss coefficients are calculated from the 6$^{th}$-order Runge-Kutta analysis, for the calculation of Eq. (S2). All other parameters are the same as in Fig. 1 in the main text.

**Note S6. Derivation of the limit cycle equation**

From $\psi_G(t) = [I(t)]^{1/2} \cdot \cos[\theta(t)] \cdot \exp[+i\phi(t)/2] \cdot \exp[i\omega(t)]$ and $\psi_L(t) = [I(t)]^{1/2} \cdot \sin[\theta(t)] \cdot \exp[-i\phi(t)/2] \cdot \exp[i\omega(t)]$ with real-valued parameters[9] $I(t)$, $\theta(t)$, $\phi(t)$, and $\omega(t)$, the derivatives of the field amplitudes $\psi_{G,L}$ are

$$\frac{d\psi_G}{dt} = \left[\frac{1}{2I} \cdot \frac{dI}{dt} - (\tan\theta) \cdot \frac{d\theta}{dt} + \frac{i}{2} \cdot \frac{d\phi}{dt} + i\frac{d\omega}{dt}\right] \cdot \psi_G$$
$$\frac{d\psi_L}{dt} = \left[\frac{1}{2I} \cdot \frac{dI}{dt} + (\cot\theta) \cdot \frac{d\theta}{dt} - \frac{i}{2} \cdot \frac{d\phi}{dt} + i\frac{d\omega}{dt}\right] \cdot \psi_L$$
(S3)

With Eq. (S3), Eq. (1) in the main text then becomes

$$\left[\frac{1}{2I} \cdot \frac{dI}{dt} - (\tan\theta) \cdot \frac{d\theta}{dt} + \frac{i}{2} \cdot \frac{d\phi}{dt} + i\frac{d\omega}{dt}\right] - \left(i\omega_0 + \gamma_G \cdot \frac{1}{1+\frac{I\cos^2\theta}{I_G}}\right) = i\kappa \cdot e^{-i\phi} \cdot \tan\theta$$

$$\left[\frac{1}{2I} \cdot \frac{dI}{dt} + (\cot\theta) \cdot \frac{d\theta}{dt} - \frac{i}{2} \cdot \frac{d\phi}{dt} + i\frac{d\omega}{dt}\right] - \left(i\omega_0 - \gamma_L \cdot \frac{1}{1+\frac{I\sin^2\theta}{I_L}}\right) = i\kappa \cdot e^{i\phi} \cdot \cot\theta$$
(S4)

Because $I(t)$, $\theta(t)$, $\phi(t)$, and $\omega(t)$ are real, Eq. (S4) can be divided into real and imaginary parts as

$$\frac{1}{2I} \cdot \frac{dI}{dt} - (\tan\theta) \cdot \frac{d\theta}{dt} - \gamma_G \cdot \frac{I_G}{I_G + I\cos^2\theta} = \kappa \cdot \tan\theta \cdot \sin\phi$$
$$\frac{1}{2I} \cdot \frac{dI}{dt} + (\cot\theta) \cdot \frac{d\theta}{dt} + \gamma_L \cdot \frac{I_L}{I_L + I\sin^2\theta} = -\kappa \cdot \cot\theta \cdot \sin\phi$$
$$\frac{1}{2} \cdot \frac{d\phi}{dt} + \frac{d\omega}{dt} - \omega_0 = \kappa \cdot \tan\theta \cdot \cos\phi$$
$$-\frac{1}{2} \cdot \frac{d\phi}{dt} + \frac{d\omega}{dt} - \omega_0 = \kappa \cdot \cot\theta \cdot \cos\phi$$
(S5)

The combination of the first and second equations in Eq. (S5) derives the first and second equations of Eq. (3) in the main text, while the combination of the third and fourth equations of Eq. (S5) derives the third equation of Eq. (3) in the main text with $d\omega/dt = \omega_0 + \kappa \cdot \cos(\phi)/\sin(2\theta)$.

# References

1. Guo, A., Salamo, G., Duchesne, D., Morandotti, R., Volatier-Ravat, M., Aimez, V., Siviloglou, G. & Christodoulides, D. Observation of PT-symmetry breaking in complex optical potentials. *Phys. Rev. Lett.* **103**, 093902 (2009).

2. Yu, S., Piao, X., Mason, D. R., In, S. & Park, N. Spatiospectral separation of exceptional points in PT-symmetric optical potentials. *Phys. Rev. A* **86**, 031802 (2012).

3. Rüter, C. E., Makris, K. G., El-Ganainy, R., Christodoulides, D. N., Segev, M. & Kip, D. Observation of parity–time symmetry in optics. *Nat. Phys.* **6**, 192-195 (2010).

4. Peng, B., Özdemir, Ş. K., Lei, F., Monifi, F., Gianfreda, M., Long, G. L., Fan, S., Nori, F., Bender, C. M. & Yang, L. Parity–time-symmetric whispering-gallery microcavities. *Nat. Phys.* **10**, 394-398 (2014).

5. Chang, L., Jiang, X., Hua, S., Yang, C., Wen, J., Jiang, L., Li, G., Wang, G. & Xiao, M. Parity–time symmetry and variable optical isolation in active–passive-coupled microresonators. *Nat. Photon.* **8**, 524-529 (2014).

6. Assawaworrarit, S., Yu, X. & Fan, S. Robust wireless power transfer using a nonlinear parity–time-symmetric circuit. *Nature* **546**, 387-390 (2017).

7. Choi, Y., Hahn, C., Yoon, J. W., Song, S. H. & Berini, P. Extremely broadband, on-chip optical nonreciprocity enabled by mimicking nonlinear anti-adiabatic quantum jumps near exceptional points. *Nat. Commun.* **8**, 14154 (2017).

8. Hassan, A. U., Zhen, B., Soljačić, M., Khajavikhan, M. & Christodoulides, D. N. Dynamically Encircling Exceptional Points: Exact Evolution and Polarization State Conversion. *Phys. Rev. Lett.* **118**, 093002 (2017).

9. Sukhorukov, A. A., Xu, Z. & Kivshar, Y. S. Nonlinear suppression of time reversals in PT-symmetric optical couplers. *Phys. Rev. A* **82**, 043818 (2010).